%% file: main-arxiv.tex
\documentclass[lettersize,journal]{IEEEtran}
\IEEEoverridecommandlockouts
\usepackage{cite}
\usepackage{amsmath,amssymb,amsfonts}
\usepackage{booktabs}
\usepackage[acronym]{glossaries}
\usepackage{siunitx}
\usepackage{subcaption}
\usepackage{hyperref}
\usepackage{xurl} 

\usepackage{pgfplots}
\usetikzlibrary{arrows.meta,calc,patterns}
\usepgfplotslibrary{groupplots}
\usetikzlibrary{calc}
\pgfplotsset{compat=1.18}

\newcommand*\myglsentry[1]{%
  \protect\ifglsused{#1}{%
    \glsentryshort{#1}%
  }{%
    \glsentrylong{#1}%
  }%
}
\loadglsentries[acronym]{./glossary}

\DeclareSIUnit{\bits}{b}
\DeclareSIUnit{\bps}{bps}
\DeclareSIUnit{\Bps}{Bps}
\DeclareSIUnit{\dBm}{dBm}

\newcommand\copyrightnotice{%
\begin{tikzpicture}[remember picture,overlay]
\node[anchor=south,yshift=20pt] at (current page.south)
  {\parbox{0.8\textwidth}{\centering\footnotesize
  This work has been submitted to the IEEE for possible publication. Copyright may be transferred without notice, after which this version may no longer be accessible.}};
\end{tikzpicture}}

\title{Disaggregated LLM Inference in Space: A Communication Feasibility Study} 

\author{{
    Arno Troch, \IEEEmembership{Graduate Student Member,~IEEE}, 
    Roger Sanchez-Vital, \IEEEmembership{Member,~IEEE},
    Seyyidahmed Lahmer,\\
    Bartek Kozicki,
    Jeroen Famaey, \IEEEmembership{Senior Member,~IEEE}, and 
    Michael Peeters, \IEEEmembership{Senior Member,~IEEE}
}

    \thanks{Arno Troch, Roger Sanchez-Vital, and Jeroen Famaey are with IDLab, Department of Computer Science, University of Antwerp--imec, Antwerp, Belgium (e-mail: arno.troch@uantwerpen.be; roger.sanchezvital@uantwerpen.be; jeroen.famaey@uantwerpen.be).}
    \thanks{Seyyidahmed Lahmer is with imec LLC Doha, Doha, Qatar (e-mail: seyyidahmed.lahmer@imec-int.com).}
    \thanks{Bartek Kozicki and Michael Peeters are with imec, Leuven, Belgium (e-mail: bartek.kozicki@imec.be; michael.peeters@imec.be).}
    \thanks{This research was funded in part by the European Union’s Horizon Europe research and innovation programme under the Marie Skłodowska-Curie grant agreement No. 101282240.}
}

\begin{document}

\maketitle

\begin{abstract}
    Orbital data centers could enable large-scale AI inference, but efficiently serving a frontier language model to many users may require several compute satellites to cooperate over optical links. While disaggregated inference, which splits large language model (LLM) serving across machines, is well established in terrestrial data centers, its viability under the unique constraints of orbital deployments remains an open question. This article investigates how link capacity and satellite separation affect two types of disaggregated LLM serving, in both compact satellite formations and distributed constellations. Prefill--decode disaggregation (PDD) migrates a large KV cache once per request, while attention--FFN disaggregation (AFD) exchanges small activations at every mixture-of-experts layer for every generated token. The two respond to distance in different ways. PDD is limited mainly by link capacity and, with fast enough links, tolerates separations of up to a thousand kilometers. AFD is limited mainly by propagation delay, which accumulates layer by layer and forces attention and expert instances within about a kilometer of each other. Based on these results, we propose a hierarchical orbital inference architecture and provide network design guidelines for future orbital data centers, including the role of ground access latency.
\end{abstract}

\copyrightnotice

\section{Introduction}
\label{sec:introduction}

Serving \glspl{LLM} requires substantial computing infrastructure. Today's frontier models run on accelerator clusters that require large amounts of power, cooling, and high-bandwidth networking. Building this infrastructure can take years, and securing grid capacity can take even longer. The International Energy Agency estimates that data centers consumed 415~TWh of electricity worldwide in 2024 and projects consumption of about 945~TWh by 2030~\cite{ieaenergy}. These demands have prompted interest in alternatives to terrestrial expansion.

One proposed direction is to deploy computing infrastructure in space. Satellites in suitable sun-synchronous orbits could access near-continuous solar energy, while optical \glspl{ISL} could connect multiple compute nodes together. Orbital deployment would avoid many terrestrial constraints, such as grid-connection queues and local zoning restrictions. Solar generation would also reduce dependence on purchased electricity. Efforts to develop orbital data centers range from research studies and proposed rack-scale systems to in-orbit demonstrations. To give a few examples, Google's Project Suncatcher investigates tightly clustered satellites equipped with \glspl{TPU} and high-capacity optical links~\cite{suncatcher}, SpaceXAI has announced a first-generation Starmind satellite design based on an optimized NVIDIA Vera Rubin NVL72 system~\cite{spacexai}, and Starcloud has flown an NVIDIA H100 GPU on Starcloud-1~\cite{starcloud}.

Even if individual satellites can host rack-sized computing infrastructure, efficiently running frontier \gls{LLM} inference services at scale may require multiple satellites to cooperate. Terrestrial inference systems typically disaggregate different stages of LLM inference across servers or racks to improve resource utilization and scale bottleneck stages independently. Two relevant approaches are \gls{PDD}~\cite{distserve,splitwise}, which separates prompt processing from token generation, and \gls{AFD}~\cite{megascale,step3}, which separates attention from expert computation. These approaches put different requirements on the underlying hardware infrastructure, leading to heterogeneous deployment and resulting in different communication patterns. Terrestrial systems can rely on physical co-location and high-capacity fiber optic networks with low inter-cluster propagation delay, whereas \glspl{ISL} must account for the separation and motion of the spacecraft.

This raises the question: \emph{which forms of inference disaggregation can an orbital network support, and how should the serving stages be distributed across satellites?} Our contributions are as follows:

\begin{itemize}
    \item We define the communication patterns of \gls{PDD} and \gls{AFD}, quantify PDD state-transfer costs using a terrestrial benchmark, and derive AFD bandwidth--distance requirements for three open-weight models and representative serving targets.
    \item We use these requirements to develop deployment guidelines and propose a hierarchical architecture for orbital disaggregated inference.
\end{itemize}

\section{Background}
\label{sec:background}

\begin{figure*}[t]
    \centering
    \includegraphics[width=0.78\textwidth]{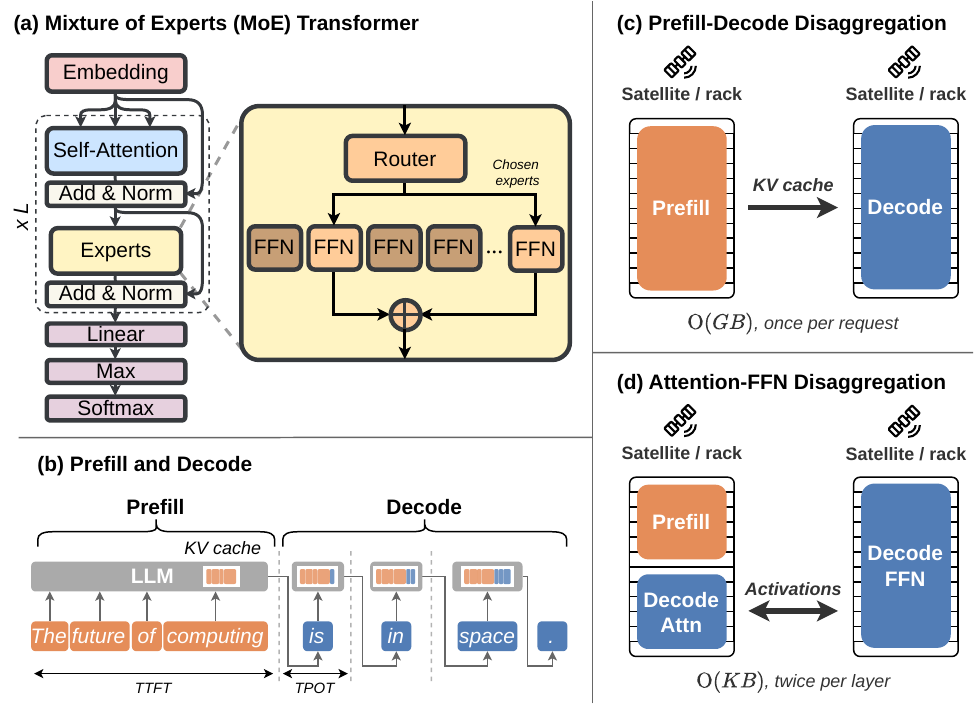}
    \caption{Overview of disaggregated \acrfull{MoE} inference. (a) A \acrshort{MoE} Transformer architecture. (b) The LLM processes the input prompt during prefill, then repeatedly generates tokens during decode. (c) PDD transfers attention state between prefill and decode instances. (d) AFD exchanges activations with the expert instance at each disaggregated MoE layer. The different communication bandwidth and latency characteristics determine how strongly inter-satellite distance affects each approach.}
    \label{fig:intro}
\end{figure*}

An \gls{LLM} represents text as tokens and processes these tokens through a sequence of Transformer layers. Each layer typically consists of an \textit{attention} block, which relates tokens to each other, and a \textit{feed-forward network} (FFN) block, which transforms the resulting representation. In a \gls{MoE} model~\cite{shazeer_2017_OutrageouslyLargeNeural}, as illustrated in Fig.~\ref{fig:intro}(a), the single \gls{FFN} component is replaced by many smaller \gls{FFN} ``experts" and an additional router component selects only a few experts for each token.

\gls{LLM} inference consists of two stages (cf. Fig.~\ref{fig:intro}(a)). During \emph{prefill}, the model processes the user prompt and produces a \textit{\gls{KV} matrix} of its tokens. This matrix is typically stored in a \emph{\gls{KV} cache} for reuse during token generation. During \emph{decode}, the model generates new tokens one at a time, conditioned on the prompt and all previously generated tokens. At each step, it reuses the cached representations and adds those of the newly processed token. This avoids recomputing the \gls{KV} representations of earlier tokens at every decoding step.

There are two important metrics to quantify the performance of prefill and decode. For the prefill stage, the \gls{TTFT} measures the wait from submitting a request to receiving its first output token. It can include queuing, request batching, prompt processing, and network delays. For the decode stage, the \gls{TPOT} describes the average interval between generated tokens. For example, a \gls{TPOT} of 10~ms corresponds to generating 100 tokens per second for one user.

Prefill and decode have very different computational characteristics. Prefill is typically \textit{compute bound}, since it can perform many computations over all the prompt tokens in parallel. Decode is often \textit{memory bound}, since each step reads model weights and the \gls{KV} cache needed for the new token. As a result, placing prefill and decode on the same hardware may cause inefficiencies, as it couples the resource allocation and batching policies of the two phases. For example, long prefill batches can delay ongoing decode steps when they share accelerator resources. Therefore, these phases can benefit from being deployed on different hardware instances.

\subsection{Disaggregated MoE Inference}

One way to address this, is \textit{prefill--decode disaggregation} (PDD), where prefill and decode stages of \gls{LLM} inference are placed on separate hardware instances, either within a single rack or across racks. This allows the two instances to scale independently. For example, decode requests from multiple prefill servers can be batched together in order to improve the hardware utilization of a decode instance. Splitwise and DistServe demonstrate this in terrestrial systems~\cite{splitwise,distserve}. After the prefill stage has completed, the \gls{KV} cache is transferred to the decode server. While the volume of the \gls{KV} cache can be large (order of gigabytes), there is only a single data transfer per request.

Another complementary approach is \textit{attention--FFN disaggregation} (AFD), where an additional type of disaggregation is performed at the decode stage by putting the attention and expert blocks on separate instances. The key idea of \gls{AFD} is to exploit the different scaling characteristics of attention and expert computation during decoding. Attention is primarily limited by memory bandwidth and capacity, whereas \gls{FFN} computation benefits from larger batches through increased weight reuse. By aggregating tokens from multiple attention instances into a shared \gls{FFN} pool, AFD increases the effective expert batch size and improves compute utilization. Systems such as MegaScale-Infer and Step-3 explore this organization and use microbatch pipelining to overlap computation and data movement~\cite{megascale,step3}. For each generated token, AFD sends intermediate activations between the attention and expert instance \textit{twice per MoE layer}. Therefore, unlike PDD, AFD transfers small data volumes, but communicates much more frequently. The following section explains how these different disaggregation types are reflected in the communication profiles between the elements of orbital data centers.

\subsection{Orbital Deployment}

\gls{LEO} keeps launch distance and ground access delay short, which makes it a natural candidate for cost-efficient orbital compute~\cite{orbital_data_centers}. Within \gls{LEO}, the main concern is continuity of power, since in most orbits a satellite spends up to 36~min of every 97~min revolution in Earth's shadow~\cite{orbital_data_centers}. A node that computes continuously must then carry a battery and extra solar array, which can account for about one third of its launched component mass~\cite{orbital_data_centers}. A dawn--dusk \gls{SSO} avoids most of this cost. Its orbital plane follows the boundary between day and night, so eclipses are absent or brief and seasonal. Most recent orbital data center proposals therefore target this orbit, including Google's Suncatcher at 650~km~\cite{suncatcher}.

This orbit is chosen for its near-continuous access to sunlight, not for proximity to users~\cite{orbital_data_centers}. Users near local sunrise or sunset lie close to the orbital plane, whereas users near local noon or midnight are about a quarter of the globe away.

Within a given orbit, satellites can be spaced differently. We consider two regimes, which we use throughout the article:
\begin{itemize}
    \item \emph{Compact formation} (\SIrange{100}{300}{m}). Google's Suncatcher describes an illustrative formation of 81 satellites within a radius of about \SI{1}{km}, with neighboring satellites roughly \SIrange{100}{300}{m} apart~\cite{suncatcher}. Short links make high optical capacity easier to achieve.
    \item \emph{Distributed constellation} (\SIrange{100}{1000}{km}). Satellites are spaced evenly along one orbital plane, so their separation is the orbit's circumference divided by the number of satellites. At \SI{650}{km}, the circumference is about \SI{44100}{km}. A separation of \SI{1000}{km} therefore corresponds to 44 satellites per plane, and \SI{100}{km} to 441. The direct optical path between neighbors is a straight line and thus slightly shorter than this arc.
\end{itemize}

The two disaggregation approaches are exposed to this inter-satellite propagation delay in different ways. In \gls{PDD}, the prefill instance sends the \gls{KV} cache to the decode instance once, after the prompt has been processed. The propagation delay is therefore incurred a single time per request, and it adds to the \gls{TTFT}. At \SI{100}{km}, this corresponds to \SI{0.33}{ms}. As explained in the previous subsection, in \gls{AFD}, the attention instance sends activations to the expert instance at every \gls{MoE} layer and must wait for the results before it can continue with the next layer. Each layer therefore incurs one full round trip, and generating a token requires one pass through all layers. The per-token propagation delay therefore depends on the inter-satellite distance and the number of layers the model has. As an example, for Qwen3-235B (94 layers) at \SI{100}{km}, this results in a total propagation delay of \SI{63}{ms} per token, before any serialization or processing. This delay adds to the \gls{TPOT} and caps token generation at roughly $1 / 63~\text{ms} \approx 16$ tokens per second per user. Additionally, compute resources remain idle during the time data is in flight, causing hardware utilization to drop.

\section{Disaggregated Inference in Space}
\label{sec:space_inference}

This section quantifies the communication requirements that \gls{PDD} and \gls{AFD} place on \glspl{ISL} in the two satellite formations introduced in Section~\ref{sec:background}. We first state our assumptions, then analyze each approach together with the ground access delay, and finally translate the results into a proposed orbital architecture.

\subsection{Assumptions}
Inspired by the Starmind satellite concept~\cite{spacexai}, we assume that each satellite can accommodate one compute rack, such as the NVIDIA Vera Rubin NVL72~\cite{rubin}, together with the required solar panels, cooling, and networking hardware. An \gls{LLM} serving instance may use one or more racks, with sufficient memory for its assigned weights and request state. We hold compute capacity fixed in order to isolate communication characteristics in different formations and with various link capacities. We leave the establishment of power, thermal, launch, or radiation feasibility of the assumed spacecraft for future work and focus solely on the communication aspects.

For AFD, we consider three open-weight models: Qwen3-235B-A22B, MiniMax-M2.5, and Kimi K3. We disaggregate 94, 62, and 92 MoE layers. For PDD, we use a separate calibration from Mooncake's Qwen3-235B-A22B-Instruct-2507 benchmark~\cite{mooncakebenchmark}.

In terms of communication, we model a direct inter-satellite optical link as a combination of propagation delay (i.e., dependent on the distance) and serialization delay (i.e., dependent on the channel capacity). To isolate the effect of propagation delay, we hold capacity fixed in each distance comparison. We exclude contention, retransmissions, endpoint overhead, and multi-hop communication. Based on Google Suncatcher's early optical link demonstrator~\cite{suncatcher}, we assume a maximum \gls{ISL} capacity of 800~Gbps. Achieving the same capacity over a longer path may be unrealistic due to a lower optical link budget.

\subsection{Communication Requirements}

PDD and AFD have different communication requirements that are inherent to their operation. We analyze how much link capacity PDD requires for KV cache transfer and how far apart attention and experts can be placed under an AFD latency budget. The two analyses use different metrics: PDD reports the fraction of service-side TTFT occupied by a \gls{KV} cache transfer, while AFD limits the additional communication time per generated token. In both cases, we consider both propagation and serialization delay.

\subsubsection{PDD}

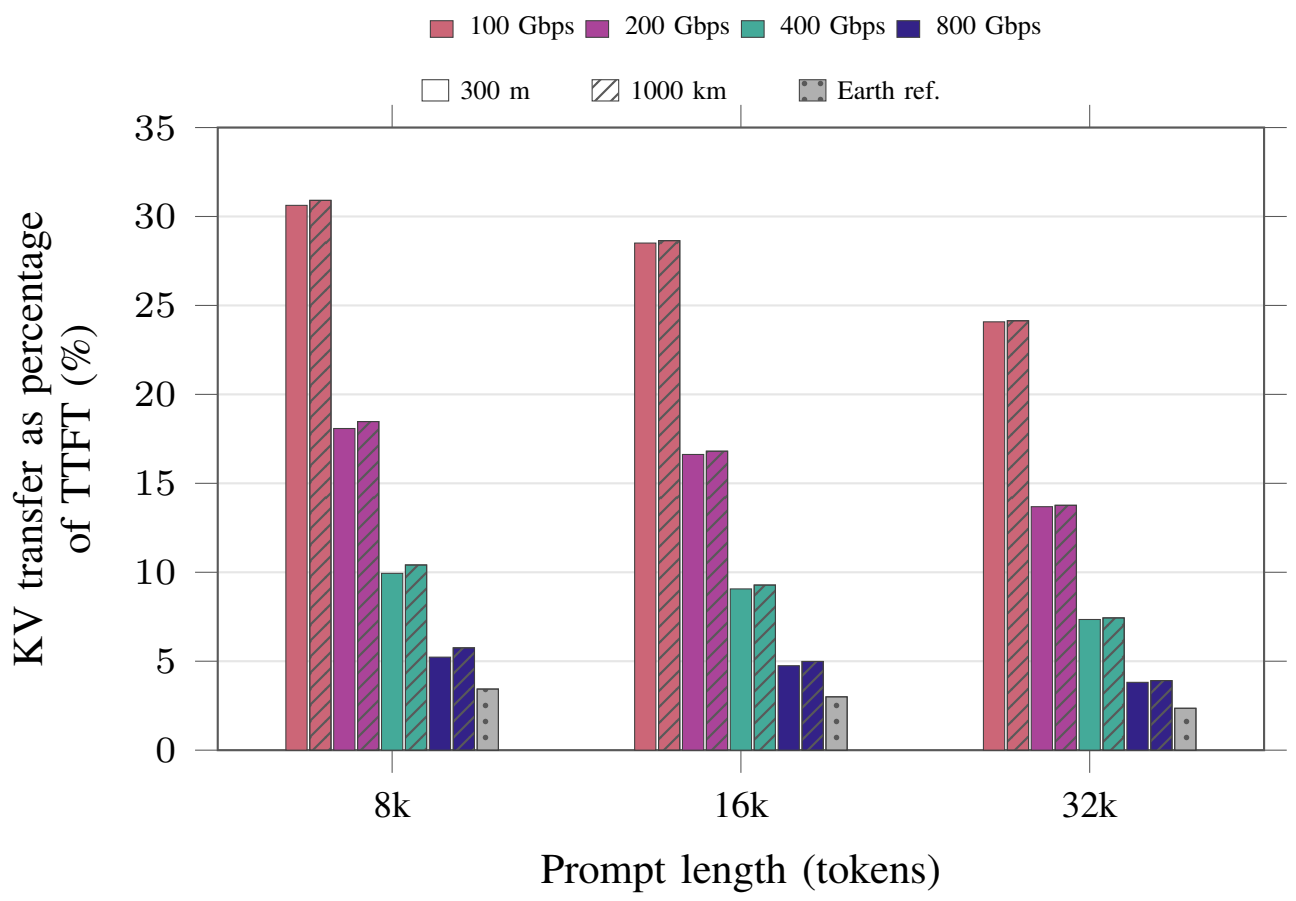
\begin{figure}[t]
    \centering
    \input{figures/pdd_results}
    \caption{Percentage of total TTFT spent on KV cache transfer time for 8k, 16k, and 32k input prompt lengths using Qwen3-235B-A22B-Instruct-2507. Gray bars represent the terrestrial reference from Mooncake~\cite{mooncakebenchmark}. The colored bars replace Mooncake's terrestrial transfer time with orbital ones at 100--800~Gbps, and retain the terrestrial computation time.}
    \label{fig:pdd}
\end{figure}

For PDD, we use Mooncake's reported TTFT measurements for \gls{KV} cache transfer volumes of 3.15, 6.31, and 12.62~GB at 8k, 16k, and 32k input prompt length in tokens, respectively~\cite{mooncakebenchmark}.

Fig.~\ref{fig:pdd} shows the percentage of TTFT spent on KV cache transfer as a function of input prompt length for different \gls{ISL} capacities and distances. The colors distinguish link capacities ranging from 100 to 800~Gbps, and hatching distinguishes satellite separations of 300~m and 1,000~km. The gray bars show the measured terrestrial reference from Mooncake. For the orbital scenarios, we replace Mooncake's measured KV cache transfer time with the estimated inter-satellite transfer time, while keeping the remaining TTFT unchanged. 

Results show that, over the considered range, link capacity matters more than inter-satellite separation. Extending the path from 300~m to 1000~km adds only 3.33~ms of propagation delay, which is small next to the time needed to transfer a multi-gigabyte cache. Increasing the capacity from 100 to 800~Gbps reduces serialization delay from about 24--31\% to 3.8--5.8\% of the modeled TTFT (i.e., up to a $6\times$ decrease). At the highest capacity, the fraction of TTFT spent on data transfer becomes comparable to the terrestrial reference, suggesting that PDD would be viable if these capacities can be reached in practical deployments.

The transfer fraction decreases with prompt length because the prefill compute overhead, which makes up most of the TTFT, grows faster than the KV cache transfer time. Therefore, in order to reduce the overhead of KV cache transfer, PDD requires sufficient capacity even when the KV cache is small. Other optimizations unrelated to network capacity could also significantly reduce transfer overhead. For example, better model architectures (e.g., reducing the required KV cache size) or system optimizations (e.g., partially overlapping prefill computation and KV cache transfer) can help to reduce the exposed delay. Each of these options requires its own assessment for orbital deployment.

\subsubsection{AFD}

\begin{figure}[t]
    \centering
    \input{figures/afd_capacity_distance}
    \caption{Combination of inter-satellite separation and capacities that satisfy the AFD communication budget (10\% of TPOT). Solid and dashed curves represent 10\% overhead for TPOT values of 10~ms and 1~ms, respectively. The shading denotes the distance range of compact formations (100--300~m) and distributed constellations (100--1000~km).}
    \label{fig:afd}
\end{figure}
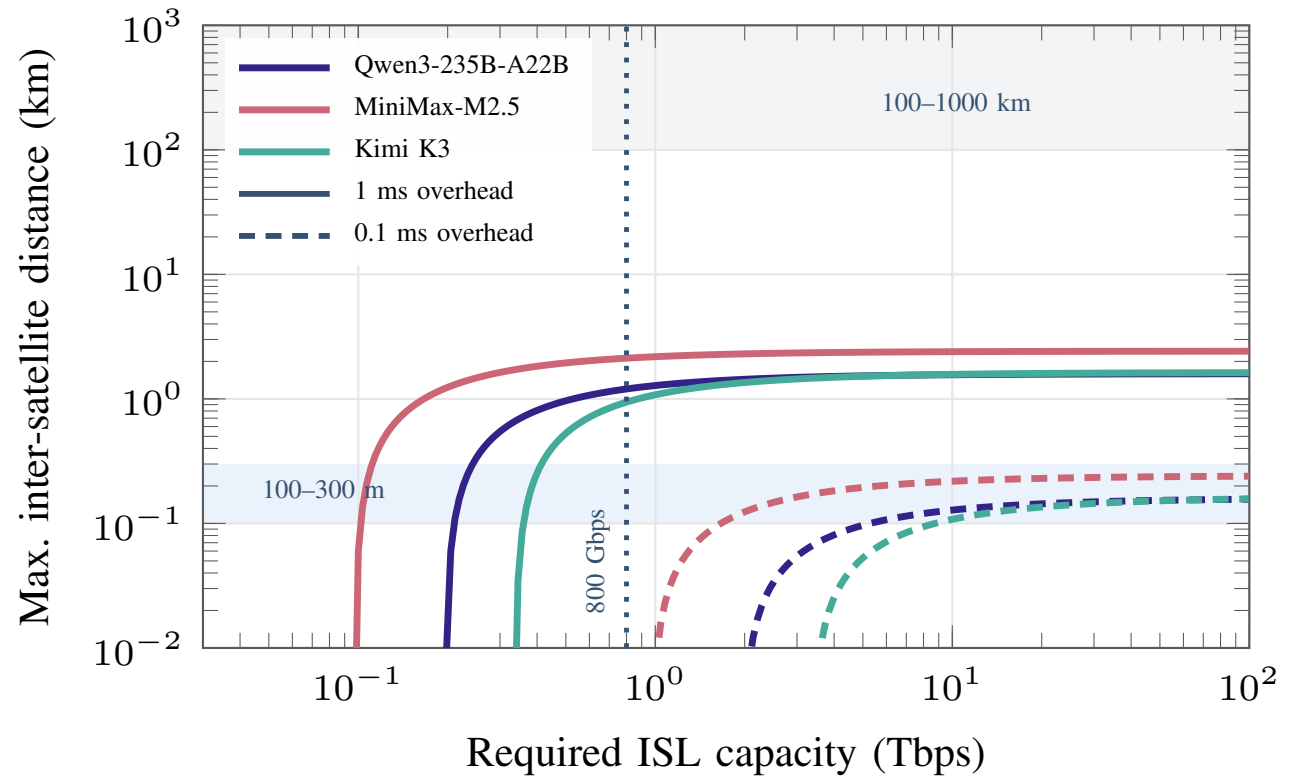

For AFD, one vector of activations in BF16 precision is sent to an expert pool and one combined result is returned. We consider a wide range of typical TPOT values of 1~ms and 10~ms, corresponding to 1000 and 100~tokens/s/user. Since TPOT includes both computation and communication, we allocate a conservative 10\% of this interval to inter-satellite activation transfers. We model the communication time as the sum of serialization and propagation delays for two transfers per disaggregated layer, assuming that these transfers are fully exposed and there is no overlap with computation. This provides a conservative communication feasibility model with respect to potential computation–communication overlap, without predicting the actual TPOT of an AFD deployment.

Fig.~\ref{fig:afd} shows the combinations of inter-satellite link capacity versus maximum inter-satellite separation that satisfy the AFD communication budget of 10\% of TPOT. Colors distinguish the different models, while solid and dashed curves represent overheads for TPOT values of 10~ms and 1~ms, respectively. Each curve indicates the maximum allowable separation for a given link capacity, or equivalently, the minimum required capacity for a given separation. Therefore, configurations below each curve satisfy the corresponding communication budget, while those above the curve exceed it.

A 10~ms reference TPOT gives a 1~ms communication overhead (i.e., 10\% of the TPOT, as defined previously). The maximum inter-satellite separation that would satisfy this overhead at \SI{800}{Gbps} is around \SIrange{1}{3}{km} for the three \gls{MoE} models considered. Furthermore, for smaller distances between 100-300~m, we notice that the required capacity remains in a respectable and realistic range from 100-400~Gbps. Therefore, a compact satellite formation would support \gls{AFD} for this specific TPOT target and overhead combination.

With a 1~ms reference TPOT (cf. dashed line in Fig.~\ref{fig:afd}), the maximum overhead is 0.1~ms. As capacity increases, serialization delay decreases, and the curves approach a distance ceiling set by propagation delay. Therefore, even higher capacity links cannot extend AFD beyond a few hundred meters under this budget, as serialization alone already exceeds the allowed overhead for every model.

\subsubsection{End-to-end delay to Earth}

Our PDD and AFD analysis covers communication within different orbital formations. However, an LLM inference user on Earth also needs to wait for the prompt to reach the cluster, and for the response to return. This access path adds to user-observed response time, while it is excluded from the service-side latency values analyzed before.

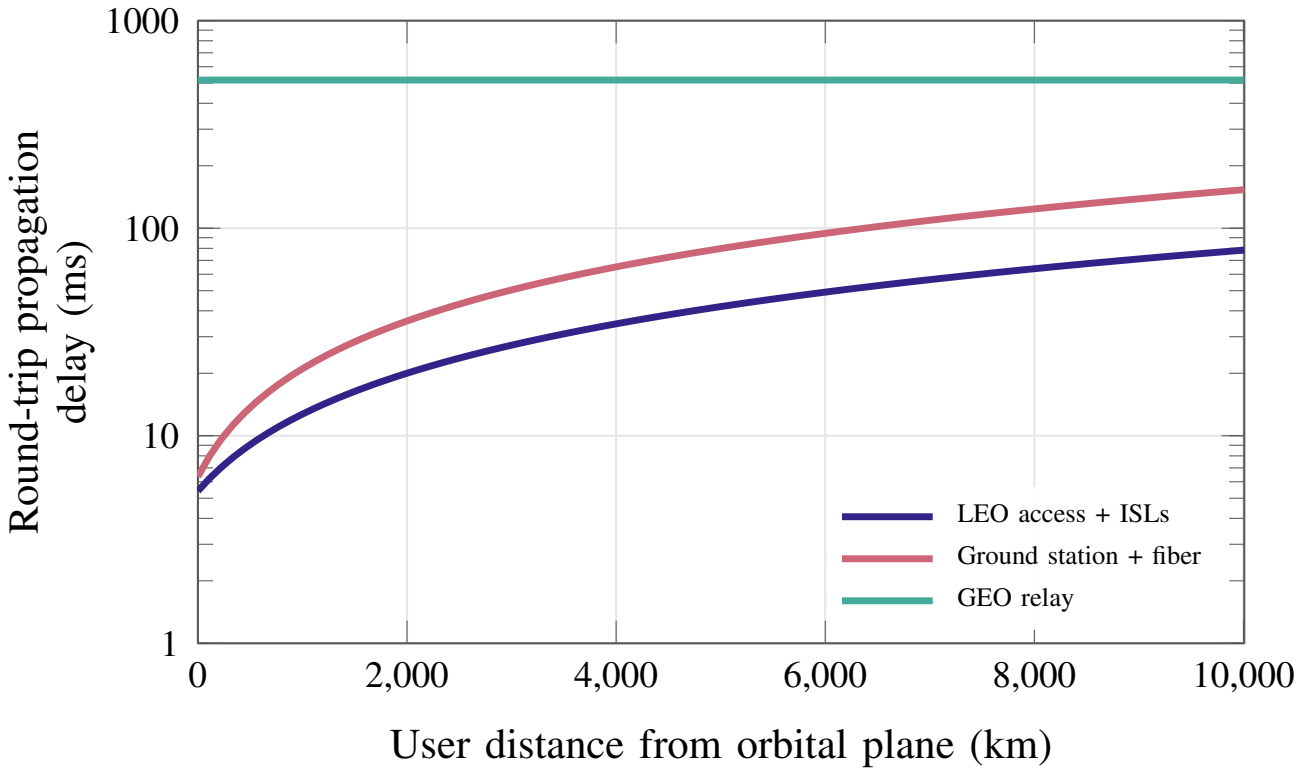
\begin{figure}[t]
    \centering
    \input{figures/e2e_ground_delay}
    \caption{Round-trip propagation delay between a ground user and a compute cluster in a 650~km dawn--dusk SSO, versus the user's surface distance from the orbital plane (10,000~km $\approx$ 90$^\circ$). Processing, queuing, and serialization are excluded.}
    \label{fig:e2e}
\end{figure}
 
The round-trip time depends on the user's position relative to the orbit and on the relay path. Fig.~\ref{fig:e2e} compares three different access paths to a cluster at 650~km. In the \textit{\gls{LEO} access path}, the user connects to a 550~km access satellite at 40$^\circ$ elevation, and traffic then follows an \gls{ISL} arc at 600~km to the cluster. In the \textit{ground station + fiber path}, the user first reaches a ground station over fiber, and the ground station then links directly to the cluster at 40$^\circ$ elevation. The fiber segment is modeled at about 7.3~$\mu$s per great-circle kilometer (group index 1.468, route factor 1.5). In the \textit{geostationary} path, each direction takes two hops of about 38,800~km. Directly below the orbital plane, the round trip is about 5.4~ms through the access constellation and 6.4~ms through fiber. At 10,000~km from the plane (roughly local noon or midnight at the equator), the modeled round trip grows to about 80~ms through the access constellation and 150~ms through fiber. The geostationary relay incurs a delay of about 520~ms at any distance. We only consider propagation delay and do not consider any processing, queuing, and routing detours that add delay in practical deployments.

Ground access delay is an important metric for the user-perceived end-to-end latency. Depending on the application that is using the LLM inference service, such delay may have a significant impact on the application’s latency requirements, regardless of the TTFT and TPOT.
For example, for interactive human chat sessions, the hundreds of milliseconds added to a response are acceptable. However, it is different for current and future real-time inference applications, such as robotics or agentic AI with repeated tool calling. More specifically, for the latter group of applications, longer tasks can amortize the initial access delay, but agentic workloads are not automatically latency tolerant. If tool execution remains on Earth, each dependent tool call may require another ground--space exchange. Placing the model, orchestration, and frequently used tools near each other can reduce these repeated exchanges. The relevant criterion is the number of sequential ground--space exchanges on the application's critical path~\cite{orbital_data_centers}.

\subsection{Proposed architecture}
\begin{figure*}[t]
    \centering
    \includegraphics[width=0.99\linewidth]{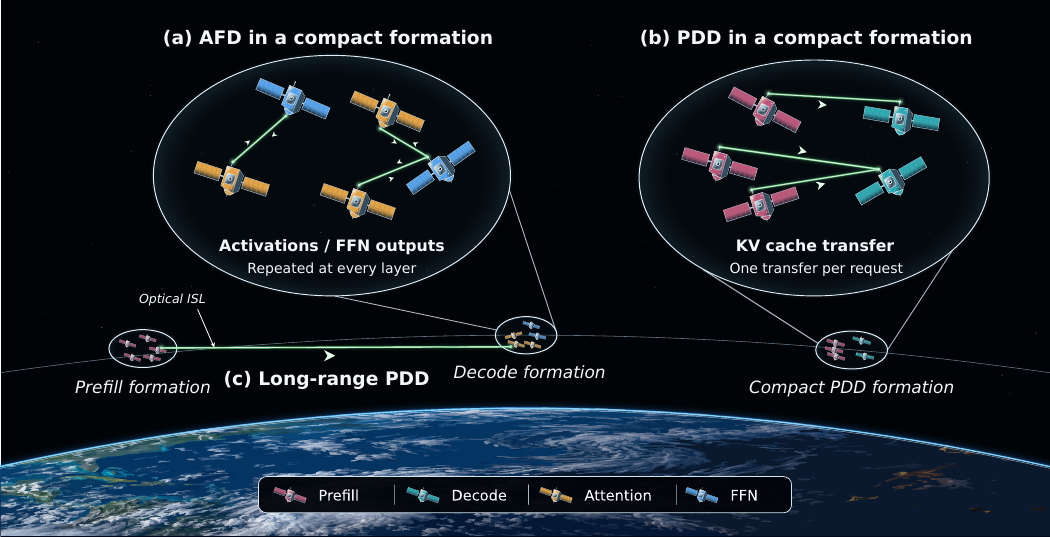}
    \caption{Proposed disaggregated inference architecture for orbital data centers. Compact satellite formations can be used for both AFD and PDD, as depicted in (a) and (b). For larger inter-satellite distances, PDD is the only feasible disaggregation technique to be deployed across multiple satellites, as illustrated in (c).}
    \label{fig:architecture}
\end{figure*}

Based on the presented results, we define a set of guidelines for future orbital system architectures for disaggregated \gls{LLM} inference. Fig.~\ref{fig:architecture} illustrates our proposed architecture. Compact groups support the frequent activation exchanges of AFD, while a network with longer \glspl{ISL} can connect prefill and decode groups for KV cache migration.

\noindent\textbf{Keep AFD instances within compact satellite formations.} Due to the frequent exchange of activations, propagation delay affects decoding performance much more than \gls{ISL} capacity. As shown in the results, for \gls{AFD} to be feasible in multi-satellite deployments, the inter-satellite distance needs to be roughly in the order of hundreds of meters.
Therefore, as depicted in Fig.~\ref{fig:architecture}, AFD is limited to deployment in compact satellite formations, where attention and expert instances that exchange activations should be placed closely together. If the TPOT target cannot be met through multi-satellite AFD, then attention and expert separation may remain within the same satellite, assuming that the satellite has enough memory to fit the model weights, activations, and KV caches of multiple requests. A shared expert pool is useful only if the increase in utilization outweighs the cost of data transfer.

\noindent\textbf{PDD may use longer range links, but it depends on total link capacity.} Our analysis shows that serialization delay is the main contributing factor to KV cache transfer time. Therefore, the main requirement over the considered distances is to have enough effective capacity. Inter-satellite distance is still an important consideration, since maintaining high-capacity links becomes more difficult as optical path loss grows. Overlapping KV cache transfers during prefill computation can reduce the exposed delay, but the full state volume still consumes bandwidth and imposes strict requirements on the total link capacity.

\section{Discussion and Open Challenges}
\label{sec:discussion}
In this section, we outline further improvements of our communication analysis, the lessons learned for orbital inference, and several open challenges.

First, our analysis provides an initial estimate of the communication requirements for disaggregated inference in orbit. Serving performance also depends on the accelerator hardware, request batching, load balancing, and how much communication can overlap with computation. Simulations driven by request traces and calibrated with hardware measurements could build on this analysis to quantify how inter-satellite communication affects serving throughput and latency under realistic loads.
 
Second, the suitability of orbital LLM inference depends on the requirements of each application. An evaluation should cover the full workflow, including where data are stored, where tools run, and how often and how much data must travel between Earth and orbit. For example, a coding agent that runs inference in orbit may repeatedly exchange repository contents, tool requests, and execution results with services on Earth. Even when the application can tolerate the propagation delay, these transfers consume link capacity and energy and can increase operating costs. Comparing latency, energy use, and cost per completed task would help determine which applications are suitable for orbital deployment and which components should run together.
 
Finally, compact formations could make high-frequency radio links worth evaluating alongside optical inter-satellite links. Optical terminals support high data rates over long distances, but require precise pointing and tracking. Over a few hundreds of meters, electronically steered \gls{RF} arrays could use wider beams, easing pointing requirements and potentially simplifying the hardware. A terrestrial demonstration achieved 221-Gbps over 500 m using technology developed for the 300 GHz band~\cite{recordoptics}. Therefore, future work should assess whether \gls{RF} links can meet the required data rates and reliability within spacecraft constraints on terminal size, power, and cost.

\section{Conclusions}
\label{sec:conclusion}

Moving LLM inference to orbit shifts the question from how much compute fits on a satellite to how the serving pipeline should be distributed across many of them. Accordingly, the communication feasibility depends on both how much data moves and how often. Under the evaluated assumptions, PDD is primarily constrained by the capacity needed to move the cached attention state, allowing greater flexibility in satellite placement when sufficient bandwidth is available. AFD introduces repeated activation transfers across model layers, making short optical paths essential for efficient decoding. High link capacity alone cannot remove the overhead of propagation delay.

With these findings, we derive a hierarchical orbital architecture: compact decode groups provide local attention–expert cooperation, while a wider network connects prefill resources and carries state migrations. Application placement must complement this internal organization, since ground access delays the first response and can have an impact on dependent tool interactions. A constant downlink propagation delay postpones the arrival of the entire response but preserves the spacing between successive tokens.

Turning these communication bounds into a practical service requires validation with measured execution times, representative workloads, scheduling policies, and realistic optical links, together with power, thermal, and spacecraft constraints. The guiding design principle is to place frequently interacting inference components close together and use longer links where communication occurs less often.

\bibliography{references}
\bibliographystyle{IEEEtran}

\vspace*{-30pt}
\begin{IEEEbiographynophoto}{Arno Troch}
    received the B.Sc. (2021) and M.Sc. (2023) in Computer Science at the University of Antwerp. He is currently a Ph.D. student at IDLab (University of Antwerp) and imec research institute, Belgium. His research interests include optical interconnects and communication architectures for distributed AI workloads.
\end{IEEEbiographynophoto}

\begin{IEEEbiographynophoto}{Roger Sanchez-Vital}
    received the Ph.D. degree in network engineering from the Universitat Polit\`ecnica de Catalunya (UPC), Barcelona, Spain, in 2025. He is currently a Marie Sk{\l}odowska-Curie Actions (MSCA) Postdoctoral Fellow at IDLab (University of Antwerp--imec), Antwerp, Belgium. His research interests include non-terrestrial networks, the Internet of Things, energy-efficient wireless communications, and the
    IEEE 802.11 standard.
\end{IEEEbiographynophoto}

\begin{IEEEbiographynophoto}{Seyyidahmed Lahmer}
    received the Ph.D. degree from the University of Padua, Italy. He is currently a researcher at imec LLC Doha, Qatar. His research interests include networked systems, machine learning, and computer architecture.
\end{IEEEbiographynophoto}

\begin{IEEEbiographynophoto}{Bartek Kozicki}
    is a Program Director at imec’s Connected Computing division, responsible for digital architectures. In his role he shapes research strategy, defines technology roadmaps, and drives technology transfer.
    Before imec, his career includes strategy and technology leadership roles at Nokia Bell Labs with focus on next-generation communication technologies, hardware accelerator design, and cloud-native RAN. Earlier, he worked as an Associate Researcher at NTT Innovation Labs on optical network systems.
    He holds a PhD from Osaka University, Japan, and has authored 50+ publications and 20+ patents.
\end{IEEEbiographynophoto}

\begin{IEEEbiographynophoto}{Jeroen Famaey}
    is currently an Associate Professor with the University of Antwerp, Belgium, and a Senior Researcher with imec, Belgium. He has co-authored over 200 papers, published in international peer-reviewed journals and conference proceedings. His current research interests include low-power distributed machine learning and wireless communications for ambient IoT devices, as well as data-driven integrated sensing and communications.
\end{IEEEbiographynophoto}

\begin{IEEEbiographynophoto}{Michael Peeters}
    is Vice President and Sector lead of Connected Computing at imec. He received his Ph.D. in Applied Physics and Photonics from the Vrije Universiteit Brussel (VUB), where he is now a Fellow. Before imec, he was CTO in Nokia’s wireline and wireless businesses and was department director at Bell Labs. He has authored more than 100 peer-reviewed publications, numerous white papers, and holds patents in the access and photonics domains.
\end{IEEEbiographynophoto}
\vfill
\end{document}

%% file: glossary.tex
\newacronym[firstplural=Radio Access Technologies (RATs)]{RAT}{RAT}{Radio Access Technology}
\newacronym{3GPP}{3GPP}{3rd Generation Partnership Program}
\newacronym{5GC}{5GC}{5G Core}
\newacronym{AI}{AI}{Artificial Intelligence}
\newacronym{AIFSN}{AIFSN}{Arbitration Inter-Frame Space Number}
\newacronym{AP}{AP}{Access Point}
\newacronym{BSS}{BSS}{Basic Service Set}
\newacronym{BvN}{BvN}{Birkhoff-von Neumann}
\newacronym{BW}{BW}{Bandwidth}
\newacronym{BWP}{BWP}{Bandwidth Part}
\newacronym{C-ITS}{C-ITS}{Cooperative \myglsentry{ITS}}
\newacronym{C-V2X}{C-V2X}{Cellular \myglsentry{V2X}}
\newacronym{C2CCC}{C2CCC}{CAR 2 CAR Communication Consortium}
\newacronym{CA}{CA}{Carrier Aggregation}
\newacronym{CAM}{CAM}{Cooperative Awareness Message}
\newacronym{CBR}{CBR}{Channel Busy Ratio}
\newacronym{CCAM}{CCAM}{Cooperative, Connected and Automated Mobility}
\newacronym{CEPT}{CEPT}{European Conference of Postal and Telecommunications}
\newacronym{CN}{CN}{Core Network}
\newacronym{CPM}{CPM}{Collective Perception Message}
\newacronym{CPO}{CPO}{Co-Packaged Optics}
\newacronym{DAG}{DAG}{Directed Acyclic Graph}
\newacronym{D2D}{D2D}{die-to-die}
\newacronym{DC}{DC}{Dual-Connectivity}
\newacronym{DCC}{DCC}{Decentralized Congestion Control}
\newacronym{DCM}{DCM}{Dual Carrier Modulation}
\newacronym{DENM}{DENM}{Decentralized Environmental Notification Message}
\newacronym{DL}{DL}{downlink}
\newacronym{DP}{DP}{data parallelism}
\newacronym{DSRC}{DSRC}{Dedicated Short Range Communications}
\newacronym{DWDM}{DWDM}{Dense Wavelength-Division Multiplexing}
\newacronym{E2E}{E2E}{End-to-End}
\newacronym{EDCA}{EDCA}{Enhanced Distributed Channel Access}
\newacronym{eNB}{eNB}{Evolved Node B}
\newacronym{EP}{EP}{expert parallelism}
\newacronym{EPC}{EPC}{Evolved Packet Core}
\newacronym{EPS}{EPS}{Electrical Packet Switching}
\newacronym{ETSI}{ETSI}{European Telecommunications Standards Institute}
\newacronym{EU}{EU}{European Union}
\newacronym{eV2X}{eV2X}{Enhanced \myglsentry{V2X}}
\newacronym{FCC}{FCC}{Federal Communications Commission}
\newacronym{FDM}{FDM}{Frequecy Division Multiplexing}
\newacronym{FFN}{FFN}{feed--forward network}
\newacronym{FR1}{FR1}{Frequency Range 1}
\newacronym{FR2}{FR2}{Frequency Range 2}
\newacronym{FSDP}{FSDP}{Fully Sharded \myglsentry{DP}}
\newacronym{FSPL}{FSPL}{Free-Space Path Loss}
\newacronym{FTR}{FTR}{Fluctuating Two-Ray}
\newacronym{FWA}{FWA}{Fixed Wireless Access}
\newacronym{GEMM}{GEMM}{General Matrix Multiplication}
\newacronym{gNB}{gNB}{G Node B}
\newacronym{GPU}{GPU}{Graphics Processing Unit}
\newacronym{GQA}{GQA}{Grouped Query Attention}
\newacronym{HARQ}{HARQ}{Hybrid Automatic Repeat reQuest}
\newacronym{HBM}{HBM}{High Bandwidth Memory}
\newacronym{HD}{HD}{High Definition}
\newacronym{IBSS}{IBSS}{Independent Basic Service Set}
\newacronym{IEEE}{IEEE}{Institute of Electrical and Electronics Engineers}
\newacronym{IoV}{IoV}{Internet of Vehicles}
\newacronym{ISAC}{ISAC}{Integrated Sensing And Communication}
\newacronym{ISL}{ISL}{inter-satellite link}
\newacronym{ITS-S}{ITS-S}{ITS-G5 Station}
\newacronym{ITS}{ITS}{Intelligent Transport Systems}
\newacronym{KPI}{KPI}{Key Performance Indicator}
\newacronym{KV}{KV}{Key--Value}
\newacronym{LEO}{LEO}{Low Earth Orbit}
\newacronym{LLM}{LLM}{Large Language Model}
\newacronym{LOS}{LOS}{Line of Sight}
\newacronym{LRR}{LRR}{Long Range Radar}
\newacronym{LTE}{LTE}{Long Term Evolution}
\newacronym{MAC}{MAC}{Medium Access Control}
\newacronym{MC}{MC}{Multi-Connectivity}
\newacronym{MCS}{MCS}{Modulation and Coding Scheme}
\newacronym{MEC}{MEC}{Multi-access Edge Computing}
\newacronym{MEMS}{MEMS}{Micro Electro-Mechanical Systems}
\newacronym{MILP}{MILP}{Mixed Integer Linear Program}
\newacronym{ML}{ML}{Machine Learning}
\newacronym{MLP}{MLP}{Multi-Layer Perceptron}
\newacronym{MME}{MME}{Mobility Management Entity}
\newacronym{MST}{MST}{Minimum Spanning Tree}
\newacronym{mmWave}{mmWave}{millimeter wave}
\newacronym{MoE}{MoE}{mixture-of-experts}
\newacronym{NAS}{NAS}{Non-Access Stratum}
\newacronym{NIC}{NIC}{Network Interface Card}
\newacronym{NLOS}{NLOS}{Non-\myglsentry{LOS}}
\newacronym{NLOSv}{NLOSv}{\myglsentry{NLOS} due to vehicles}
\newacronym{NR-U}{NR-U}{\myglsentry{NR} Unlicensed}
\newacronym{NR}{NR}{New Radio}
\newacronym{NSA}{NSA}{Non-Standalone}
\newacronym{OBU}{OBU}{On-Board Unit}
\newacronym{OCB}{OCB}{Outside the Context of a Basic Service Set}
\newacronym{OCS}{OCS}{Optical Circuit Switching}
\newacronym{OFDM}{OFDM}{Orthogonal Frequency-Division Multiplexing}
\newacronym{OFDMA}{OFDMA}{Orthogonal Frequency-Division Multiple Access}
\newacronym{OSI}{OSI}{Open Systems Interconnection}
\newacronym{P2P}{P2P}{Pedestrian-to-Pedestrian}
\newacronym{PDCP}{PDCP}{Packet Data Convergence Protocol}
\newacronym{PDD}{PDD}{prefill--decode disaggregation}
\newacronym{PDN}{PDN}{Packet Data Network}
\newacronym{PGW}{PGW}{\myglsentry{PDN} Gateway}
\newacronym{PMN}{PMN}{Perceptive Mobile Network}
\newacronym{PP}{PP}{pipeline parallelism}
\newacronym{ProSe}{ProSe}{Proximity Services}
\newacronym{PSFCH}{PSFCH}{Physical Sidelink Feedback Channel}
\newacronym{QAM}{QAM}{Quadrature Amplitude Modulation}
\newacronym{QoS}{QoS}{Quality of Service}
\newacronym{R2R}{R2R}{\myglsentry{RSU}-to-\myglsentry{RSU}}
\newacronym{RAN}{RAN}{Radio Access Network}
\newacronym{RB}{RB}{Resource Block}
\newacronym{RDCN}{RDCN}{Reconfigurable \myglsentry{DCN}}
\newacronym{RF}{RF}{radio frequency}
\newacronym{RLC}{RLC}{Radio Link Control}
\newacronym{RRC}{RRC}{Radio Resource Control}
\newacronym{RSU}{RSU}{Road Side Unit}
\newacronym{SA}{SA}{Standalone}
\newacronym{SAE}{SAE}{Society of Automotive Engineers}
\newacronym{SB-SPS}{SB-SPS}{Sensing-Based Semi-Persistent Scheduling}
\newacronym{SCH}{SCH}{Secondary Cell Handover}
\newacronym{SCM}{SCM}{Statistical Channel Model}
\newacronym{SCS}{SCS}{Subcarrier Spacing}
\newacronym{SDN}{SDN}{Software-Defined Networking}
\newacronym{SGW}{SGW}{Serving Gateway}
\newacronym{SINR}{SINR}{Signal to Interference plus Noise Ratio}
\newacronym{SL}{SL}{Sidelink}
\newacronym{SNR}{SNR}{Signal to Noise Ratio}
\newacronym{SRR}{SRR}{Short Range Radar}
\newacronym{SSO}{SSO}{sun-synchronous orbit}
\newacronym{STA}{STA}{Station}
\newacronym{TDC}{TDC}{Transmit Datarate Control}
\newacronym{TDD}{TDD}{Time Division Duplex}
\newacronym{TDM}{TDM}{Time Division Multiplexing}
\newacronym{TM}{TM}{Traffic Matrix}
\newacronym{ToR}{ToR}{Top-of-Rack}
\newacronym{TP}{TP}{tensor parallelism}
\newacronym{TPC}{TPC}{Transmit Power Control}
\newacronym{TPOT}{TPOT}{time per output token}
\newacronym{TPU}{TPU}{Tensor Processing Unit}
\newacronym{TR}{TR}{Technical Report}
\newacronym{TRC}{TRC}{Transmit Rate Control}
\newacronym{TS}{TS}{Technical Specification}
\newacronym{TTFT}{TTFT}{time to first token}
\newacronym{TXOP}{TXOP}{Transmission Opportunity}
\newacronym{UDP}{UDP}{User Datagram Protocol}
\newacronym{UE}{UE}{User Equipment}
\newacronym{UL}{UL}{uplink}
\newacronym{UPA}{UPA}{Uniform Planar Array}
\newacronym{URLLC}{URLLC}{Ultra-Reliable and Low-Latency Communications}
\newacronym{V2D}{V2D}{Vehicle to Device}
\newacronym{V2I}{V2I}{Vehicle-to-Infrastructure}
\newacronym{V2N}{V2N}{Vehicle-to-Network}
\newacronym{V2P}{V2P}{Vehicle-to-Pedestrian}
\newacronym{V2R}{V2R}{Vehicle-to-\myglsentry{RSU}}
\newacronym{V2V}{V2V}{Vehicle-to-Vehicle}
\newacronym{V2X}{V2X}{Vehicle-to-Everything}
\newacronym{VANET}{VANET}{Vehicular Ad Hoc Network}
\newacronym{VLB}{VLB}{Valiant Load Balancing}
\newacronym{WI}{WI}{Work Item}
\newacronym{WLAN}{WLAN}{Wireless Local Area Network}
\newacronym{WLO}{WLO}{Wafer-Level Optics}
\newacronym{AFD}{AFD}{attention--\myglsentry{FFN} disaggregation}

%% file: figures/pdd_results.tex
\begin{tikzpicture}
\def\pddBarWidth{4pt}
\def\pddBarGap{0.5pt} 

\definecolor{pdd100}{HTML}{CC6677}
\definecolor{pdd200}{HTML}{AA4499}
\definecolor{pdd400}{HTML}{44AA99}
\definecolor{pdd800}{HTML}{332288}
\definecolor{pddEarth}{HTML}{B0B0B0}
\begin{axis}[
 name=pddPlot,
 width=.96\columnwidth,height=5.7cm,scale only axis=false,
 tick label style={font=\scriptsize},label style={font=\footnotesize},
 ylabel style={align=center},
 ymajorgrids=true,grid style={black!10,thin},
 axis line style={black!65,thin},tick style={black!65},
 tick align=outside,minor tick num=0,
 xmin=-.5,xmax=2.5,ymin=0,ymax=35,
 xtick={0,1,2},xticklabels={8k,16k,32k},
 ytick={0,5,10,15,20,25,30,35},
 xlabel={Prompt length (tokens)},
 ylabel={KV transfer as percentage\\of TTFT (\%)},
 ybar,bar width=\pddBarWidth,clip=false,
 legend columns=4,
 legend image code/.code={\draw[#1] (0cm,-.055cm) rectangle (.15cm,.055cm);},
 legend style={
   at={(0.5,1.11)},anchor=south,
   font=\tiny,draw=none,fill=white,
   inner xsep=0pt,inner ysep=1pt,row sep=1pt,
   /tikz/every odd column/.append style={column sep=1pt},
 },
 legend cell align=left,
 /pgfplots/pddshift/.style={bar shift={#1*(\pddBarWidth+\pddBarGap)}},
 /pgfplots/pddbar/.style={draw=black!75,line width=.2pt,forget plot},
 /pgfplots/pddfar/.style={postaction={pattern=north east lines,pattern color=black!65}},
]
\addplot+[pddbar,fill=pdd100,pddshift=-4] coordinates {
(0,30.6260946027) (1,28.5067689604) (2,24.0784734350)
};
\addplot+[pddbar,fill=pdd100,pddshift=-3,pddfar] coordinates {
(0,30.9061074796) (1,28.6411458689) (2,24.1388054518)
};
\addplot+[pddbar,fill=pdd200,pddshift=-2] coordinates {
(0,18.0820027477) (1,16.6227062750) (2,13.6870651958)
};
\addplot+[pddbar,fill=pdd200,pddshift=-1,pddfar] coordinates {
(0,18.4721490726) (1,16.8054130776) (2,13.7650342944)
};
\addplot+[pddbar,fill=pdd400,pddshift=0] coordinates {
(0,9.9397158901) (1,9.0647897933) (2,7.3462909859)
};
\addplot+[pddbar,fill=pdd400,pddshift=1,pddfar] coordinates {
(0,10.4110511755) (1,9.2820784323) (2,7.4361305311)
};
\addplot+[pddbar,fill=pdd800,pddshift=2] coordinates {
(0,5.2298488134) (1,4.7476095420) (2,3.8132247952)
};
\addplot+[pddbar,fill=pdd800,pddshift=3,pddfar] coordinates {
(0,5.7516290502) (1,4.9859925921) (2,3.9100429048)
};
\addplot+[pddbar,fill=pddEarth,pddshift=4,postaction={pattern=dots,pattern color=black!65}] coordinates {
(0,3.4357343438) (1,2.9988890166) (2,2.3551135841)
};

\addlegendimage{fill=pdd100,draw=black!75,line width=.2pt}
\addlegendentry{100 Gbps}
\addlegendimage{fill=pdd200,draw=black!75,line width=.2pt}
\addlegendentry{200 Gbps}
\addlegendimage{fill=pdd400,draw=black!75,line width=.2pt}
\addlegendentry{400 Gbps}
\addlegendimage{fill=pdd800,draw=black!75,line width=.2pt}
\addlegendentry{800 Gbps}
\end{axis}
\begin{scope}[shift={([yshift=7pt]pddPlot.north)}]
 \draw[fill=white,draw=black!75,line width=.2pt] (-60pt,-2pt) rectangle (-55pt,2pt);
 \node[anchor=west,font=\tiny,inner sep=0pt] at (-53pt,0) {300 m};
 \draw[fill=white,draw=black!75,line width=.2pt,postaction={pattern=north east lines,pattern color=black!65}] (-28pt,-2pt) rectangle (-23pt,2pt);
 \node[anchor=west,font=\tiny,inner sep=0pt] at (-21pt,0) {1000 km};
 \draw[fill=pddEarth,draw=black!75,line width=.2pt,postaction={pattern=dots,pattern color=black!65}] (11pt,-2pt) rectangle (16pt,2pt);
 \node[anchor=west,font=\tiny,inner sep=0pt] at (18pt,0) {Earth ref.};
\end{scope}
\end{tikzpicture}

%% file: figures/afd_capacity_distance.tex
\begin{tikzpicture}
\definecolor{afdNavy}{HTML}{355070}
\definecolor{afdQwen}{HTML}{332288}
\definecolor{afdMini}{HTML}{CC6677}
\definecolor{afdKimi}{HTML}{44AA99}
\definecolor{afdNear}{HTML}{EAF2FB}
\definecolor{afdFar}{HTML}{F3F4F6}

\begin{loglogaxis}[
 width=.96\columnwidth,height=5.7cm,scale only axis=false,
 tick label style={font=\scriptsize},label style={font=\footnotesize},
 grid=major,grid style={black!10,thin},
 axis line style={black!65,thin},tick style={black!65},
 minor tick num=0,
 legend style={
   font=\tiny,
   draw=none,
   fill=white,
   fill opacity=.94,
   text opacity=1,
   inner xsep=2pt,
   inner ysep=1pt,
   row sep=-1pt,
   /tikz/every odd column/.append style={column sep=2pt}
 },
 legend cell align=left,
 line join=round,unbounded coords=jump,
 set layers=standard,
 xmin=.03,xmax=100,ymin=.01,ymax=1000,
 xlabel={Required ISL capacity (Tbps)},
 ylabel={Max. inter-satellite distance (km)},
 legend style={at={(0.02,0.98)},anchor=north west,
   font=\tiny,draw=none,fill=white,fill opacity=.94,text opacity=1,
   inner xsep=2pt,inner ysep=1pt,row sep=-1pt,
   /tikz/every odd column/.append style={column sep=2pt}},
]
\begin{pgfonlayer}{axis background}
 \fill[afdNear] (axis cs:.03,.1) rectangle (axis cs:100,.3);
 \fill[afdFar] (axis cs:.03,100) rectangle (axis cs:100,1000);
\end{pgfonlayer}
\addplot+[no marks,color=afdQwen,line width=1.3pt,domain=.03:100,samples=250]
 {max(0.001,299792.458*(.001/(2*94)-256*4096/(x*1e12)))};
\addlegendentry{Qwen3-235B-A22B}
\addplot+[no marks,color=afdQwen,line width=1.3pt,densely dashed,domain=.03:100,samples=250,forget plot]
 {max(0.001,299792.458*(.0001/(2*94)-256*4096/(x*1e12)))};
\addplot+[no marks,color=afdMini,line width=1.3pt,domain=.03:100,samples=250]
 {max(0.001,299792.458*(.001/(2*62)-256*3072/(x*1e12)))};
\addlegendentry{MiniMax-M2.5}
\addplot+[no marks,color=afdMini,line width=1.3pt,densely dashed,domain=.03:100,samples=250,forget plot]
 {max(0.001,299792.458*(.0001/(2*62)-256*3072/(x*1e12)))};
\addplot+[no marks,color=afdKimi,line width=1.3pt,domain=.03:100,samples=250]
 {max(0.001,299792.458*(.001/(2*92)-256*7168/(x*1e12)))};
\addlegendentry{Kimi K3}
\addplot+[no marks,color=afdKimi,line width=1.3pt,densely dashed,domain=.03:100,samples=250,forget plot]
 {max(0.001,299792.458*(.0001/(2*92)-256*7168/(x*1e12)))};
\addplot+[no marks,color=afdNavy,dotted,line width=.9pt,forget plot] coordinates {(.8,.01) (.8,1000)};
\node[font=\tiny,rotate=90,anchor=south,text=afdNavy,inner sep=1pt] at (axis cs:.74,.05) {800 Gbps};
\node[font=\tiny,text=afdNavy,anchor=west] at (axis cs:.041,.19) {100--300 m};
\node[font=\tiny,text=afdNavy,anchor=west] at (axis cs:5,240) {100--1000 km};
\addlegendimage{afdNavy,solid,line width=1.1pt}
\addlegendentry{1 ms overhead}
\addlegendimage{afdNavy,densely dashed,line width=1.1pt}
\addlegendentry{0.1 ms overhead}
\end{loglogaxis}
\end{tikzpicture}

%% file: figures/e2e_ground_delay.tex
\begin{tikzpicture}
\definecolor{e2eLEO}{HTML}{332288}
\definecolor{e2eFiber}{HTML}{CC6677}
\definecolor{e2eGEO}{HTML}{44AA99}

\def\eEarthRadius{6.371}
\def\eLightSpeed{299.792458}
\def\eElevation{40}
\def\eAccessAltitude{0.550}
\def\eISLAltitude{0.600}
\def\eComputeAltitude{0.650}
\def\eFiberIndex{1.468}
\def\eFiberRoute{1.5}
\def\eGEOAltitude{35.786}
\def\eGEOSlantExtra{3}

\pgfmathsetmacro{\eAccessSlant}{\eEarthRadius*(sqrt((1+\eAccessAltitude/\eEarthRadius)^2-cos(\eElevation)^2)-sin(\eElevation))}
\pgfmathsetmacro{\eComputeSlant}{\eEarthRadius*(sqrt((1+\eComputeAltitude/\eEarthRadius)^2-cos(\eElevation)^2)-sin(\eElevation))}
\pgfmathsetmacro{\eRTTScale}{2000/\eLightSpeed}
\pgfmathsetmacro{\eGEODelay}{2*\eRTTScale*(\eGEOAltitude+\eGEOSlantExtra)}
\pgfmathsetmacro{\eXmax}{pi/2*\eEarthRadius}

\begin{semilogyaxis}[
 width=.96\columnwidth,height=5.7cm,scale only axis=false,
 tick label style={font=\scriptsize},label style={font=\footnotesize},
 xlabel style={align=center},ylabel style={align=center},ylabel shift=-5pt,
 grid=major,grid style={black!10,thin},
 axis line style={black!65,thin},tick style={black!65},
 minor tick num=0,
 minor ytick={},
 legend style={
   at={(0.98,0.02)},anchor=south east,
   font=\tiny,draw=none,fill=white,fill opacity=.94,text opacity=1,
   inner xsep=2pt,inner ysep=1pt,row sep=-1pt,
   /tikz/every odd column/.append style={column sep=2pt}
 },
 legend cell align=left,
 line join=round,unbounded coords=jump,
 xmin=0,xmax=\eXmax,ymin=1,ymax=1000,
 xtick={0,2,4,6,8,10},
 xticklabels={0,{2,000},{4,000},{6,000},{8,000},{10,000}},
 ytick={1,10,100,1000},yticklabels={1,10,100,1000},
 xlabel={User distance from orbital plane (km)},
 ylabel={Round-trip propagation\\delay (ms)},
]
\addplot+[no marks,color=e2eLEO,line width=1.3pt,domain=0:\eXmax,samples=91]
 {\eRTTScale*\eAccessSlant+(\eRTTScale*(\eEarthRadius+\eISLAltitude))*x/\eEarthRadius};
\addlegendentry{LEO access + ISLs}
\addplot+[no marks,color=e2eFiber,line width=1.3pt,domain=0:\eXmax,samples=91]
 {\eRTTScale*\eComputeSlant+(\eRTTScale*\eFiberRoute*\eFiberIndex)*x};
\addlegendentry{Ground station + fiber}
\addplot+[no marks,color=e2eGEO,line width=1.3pt,domain=0:\eXmax,samples=2]
 {\eGEODelay};
\addlegendentry{GEO relay}
\end{semilogyaxis}
\end{tikzpicture}